\documentclass[a4paper,fleqn]{cas-dc}
\renewcommand{\printorcid}{}

\usepackage{apacite}
\usepackage[round]{natbib} 

\usepackage{breakcites}

\usepackage{float}
\usepackage[section]{placeins}

\restylefloat{table}
\usepackage{diagbox}
\usepackage{adjustbox}

\usepackage{textgreek}

\def\tsc#1{\csdef{#1}{\textsc{\lowercase{#1}}\xspace}}
\tsc{WGM}
\tsc{QE}
\tsc{EP}
\tsc{PMS}

\tsc{BEC}
\tsc{DE}

\begin{document}
\let\WriteBookmarks\relax
\def\floatpagepagefraction{1}
\def\textpagefraction{.001}
\shorttitle{Predicting market inflation expectations with news topics and sentiment}
\shortauthors{Author1 et~al} 

\pagenumbering{arabic}

\title [mode = title]{Predicting market inflation expectations with news topics and sentiment}                      

\author[1]{Sonja Tilly}

\cormark[1]

\ead{sonja.tilly.19@ucl.ac.uk}

\cortext[cor1]{Corresponding author}

\address[1]{UCL, Computer Science Dep, 66 - 72 Gower St, Bloomsbury, WC1E 6EA London, UK}

\author[1,2]{Giacomo Livan}
\address[2]{Systemic Risk Centre, London School of Economics and Political Science, London, WC2A 2AE, UK}
\ead{g.livan@ucl.ac.uk}

\nonumnote{Abbreviations. GDELT: Global Database of Events, Language and Tone; GKG: Global Knowledge Graph; GCAM: Content Analysis Measure Systems; CNT: Conviction Narrative Theory; BEIR: breakeven inflation rate; PLS: partial least squares; GGC: graphical granger causality}

\begin{abstract}
This study presents a novel approach to incorporating news topics and their associated sentiment into predictions of breakeven inflation rate (BEIR) movements for eight countries with mature bond markets. We calibrate five classes of machine learning models including narrative-based features for each country, and find that they generally outperform corresponding benchmarks that do not include such features. We find Logistic Regression and XGBoost classifiers to deliver the best performance across countries. We complement these results with a feature importance analysis, showing that economic and financial topics are the key performance drivers in our predictions, with additional contributions from topics related to health and government. We examine cross-country spillover effects of news narrative on BEIR via Graphical Granger Causality and confirm their existence for the US and Germany, while five other countries considered in our study are only influenced by local narrative.

\end{abstract}

\begin{keywords}
news sentiment \sep time series forecasting \sep big data \sep machine learning \sep knowledge graph
\end{keywords}

\maketitle

\section{Introduction}

Does news narrative impact market-based inflation expectations? Market participants follow daily news reporting -- which can be positive, neutral or negative -- and incorporate it into their views and expectations accordingly. Theories on narrative suggest that investors' behavior is influenced by news flow and its sentiment fluctuations. 

Dating back to Keynes' classic work~\citep{keynes1936general}, movement in financial markets and the economy have been considered as partially driven by human impulses and emotions  -- referred to as ``animal spirits''. With the advent of mass media, such factors have been increasingly synthesised into societal narratives spread by news sources. Indeed, Shiller suggested that viral narratives have a direct causal effect on economic activity~\citep{shiller2017narrative}. More recently, Conviction Narrative Theory (CNT) has been proposed to illustrate how changes in narrative are a precursor to changes in financial markets and the economy~\citep{tuckett2014bringing}.

There is a growing body of research exploring the role of news narrative and its implications on financial markets and the economy. In particular, developments in natural language processing and the ever-increasing ability to process large volumes of data have made it possible to operationalize the concept of narrative, and to quantitatively link it with economic fundamentals~\citep{larsen2019business} and economic fluctuations~\citep{kalamara2020making}.

The above research agenda has been applied successfully to a variety of fields \citep{rousidis2020social}, making it possible to formulate increasingly accurate predictions of financial markets \citep{allen2019daily, baker2020unprecedented} or for macroeconomic indicators \citep{ardia2019questioning, elshendy2017big, tilly2021macroeconomic, tilly2021knowledge}.

Our study contributes to existing research by incorporating themes and sentiment from global newspaper articles into predictions of breakeven inflation rates (BEIR) movements using data from the Global Database of Events, Language and Tone (GDELT) \citep{leetaru_the_nodate}. GDELT data has -- to the best of our knowledge -- not yet been applied to this field of research.
The majority of existing publications focuses on one country, mainly the US. Our analysis covers eight diverse economies with mature bond markets around the world, demonstrating the global breadth of our approach. 
A feature importance analysis enhances the interpretability of our results, showing that economic and financial topics are the key performance drivers in our predictions. We examine cross-country spillover effects of news narrative on BEIR via Graphical Granger Causality modelling and confirm their existence for the US and Germany, while five other countries considered are only influenced by local narrative.

\section{Literature review}

This section addresses a selection of existing literature on the impact of news narrative on inflation expectations.

The majority of publications in this field attempt to predict consumer inflation expectations as quantified by surveys. These works utilise predominantly regression frameworks to model monthly frequency data. The following studies use the University of Michigan Survey of Consumers as proxy for inflation expectations, with an overarching agreement that news narrative triggers a change in inflation expectations.

Doms and Morin~\citep{doms2004consumer} extract measures of tone and volume of economic reporting employing The Economist's recession index. They show that US consumers update their expectations more frequently during periods of high news coverage, with their findings supporting the stickiness of consumer expectations theory as set out by Sims~\citep{sims2003implications}, whose theory explains why the tone and volume of economic reporting impact consumer sentiment beyond the economic information contained in the reporting. 

A study by Pfajfar and Santoro~\citep{pfajfar2013news} examines the relationship between newspaper coverage on inflation and US households' inflation expectations. They test an epidemiological mechanism of expectation formation theory~\citep{mankiw2003sticky, carroll2003macroeconomic}, according to which consumers update their expectations from the media, which are assumed to transmit professional forecasters' projections. The authors find a disconnect between news on inflation, consumers' frequency of expectation updating, and the accuracy of their expectations, with consumers being more receptive to negative news. 

Similarly, work by Dr{\"a}ger and Lamla~\citep{drager2017imperfect} explores the factors driving an adjustment in US consumers' inflation expectations. The authors find that news narratives lead to an adjustment in expectations, with individual expectations becoming more precise once adjusted, supporting theories of imperfect information~\citep{woodford2001imperfect,sims2003implications,mankiw2003sticky}.

Larsen \emph{et al.}~\citep{larsen2021news} examine the role of the media in consumers' inflation expectations formation process. Findings show that the news topics covered by newspapers have predictive power for the US consumer price index and inflation expectations. The extent of consumer information rigidity changes over time given varying levels of relevant news coverage, which leads consumers to update their expectations.

D'Acunto \emph{et al.}~\citep{d2019exposure} examine US inflation expectations specific to individual consumers based on grocery shopping in the US and collect this data via surveys. They highlight that consumers base their inflation expectations on price changes (that are regarded as ``news'') with a higher sensitivity to price changes in goods that they buy frequently  that they encounter in their everyday lives.

Approaching the topic from a different angle, Mazumder \citep{mazumder2021reaction} focuses on the amount of ``Fed'' mentions in US newspapers and finds evidence that such mentions improve forecasts of US inflation expectations.

Studies on the impact of news narrative on European inflation expectations reach similar conclusions compared to research focusing on the US. For instance, a paper by Jansen and Neuenkirch~\citep{jansen2017news} collects survey data from Dutch households to examine the relationship of inflation perceptions and newspaper consumption. The authors find that perceptions of recent price changes are an important driver of the accuracy of next-year inflation expectations. These perceptions become more accurate with increased newspaper consumption.

Lamla and Lein~\citep{lamla2014role} analyse the impact of media coverage of inflation related topics on German consumers' inflation expectation from the EU Business and Consumer Survey. Results suggest that both the volume and tone related to news about inflation play a role in consumers' inflation expectation formation.

While the majority of research quantifies inflation expectations using monthly frequency consumer survey data, a few studies employ market-based measures of inflation expectations such as inflation breakeven rates or interest rate swaps, suggesting that news narrative and sentiment have a causal relationship with the economy and financial markets.
A paper by Bauer~\citep{bauer2014inflation} suggests that market-based measures of inflation expectations in the US -- defined by inflation breakeven rates and interest rate swaps -- change in response to macroeconomic news.
Bybee \emph{et al.}~\citep{bybee2020structure} leverage textual analysis of US business news to extract topic-based news attention indicators. The authors use these indicators to model US interest rate swap prices using rolling 1,000 day lasso regressions. An analysis of the model coefficients suggests that over time and depending on the location and type of market friction, different topics drive model performance.
Research by Kabiri \emph{et al.}~\citep{kabiri2021role} investigates the impact of news sentiment on the economy and financial markets in the US. The authors demonstrate that sentiment shocks have significant effects on indicators of economic activity as well as on market-based indicators such as the S\&P 500 index, credit spreads, terms spreads, interest rates and prices.

\subsection{Hypotheses formulation}

In this study, we attempt to forecast short-term movements in 10 year BEIR for eight countries, leveraging data from the Global Database of Events, Language and Tone (GDELT). Specifically, we address two main research questions, i.e., (1) whether themes from GDELT and their associated average tone have predictive power in relation to short-term movements in BEIR, and (2) whether there are cross-country spillover effects from news narrative on BEIR. Accordingly, we formulate the following hypotheses:
\begin{itemize}
    \item $H_1$: GDELT themes and their associated average tone improve predictions of short-term movements in BEIR.
    \item $H_2$: There are cross-country spillover effects from news narrative on BEIR.
\end{itemize}

By tackling the above questions, we advance the literature on the prediction of market-based inflation expectations in at least three ways. First, we demonstrate how themes and their associated sentiment from newspaper narrative can be incorporated into machine learning models to improve predictions of breakeven inflation rates (BEIR) for a range of diverse countries. Second, we contribute to the literature on forecasting inflation expectations by proposing a data-driven methodology to operationalise concepts such as information extraction, feature generation and predictive modelling. We complement our results with a feature importance analysis showing that economy-related themes have the strongest predictive power. Third, we contribute to research on the cross-country impact of news narrative on BEIR leveraging Graphical Granger Causality (GGC) and graph analytics.

\section{Data and methods}
This section introduces GDELT as data source and provides details on the variables used in this study.

The GDELT Project is a collaboration of different bodies at Google, the Yahoo! Fellowship at Georgetown University and several large news archives such as Lexis Nexis and JSTOR. The project monitors world media from a multitude of angles, identifying and extracting entities such as themes, emotions, locations and events. GDELT version two incorporates real-time translation from 65 languages and is updated every 15 minutes \citep{leetaru_the_nodate}. It is a public data set accessible on the Google Cloud Platform, comprising roughly 12 terabytes of data.

The GDELT Global Knowledge Graph (GKG) is backed by a software that scans global newspaper articles in real-time to extract entities such as persons, organizations, locations, dates, themes and emotions~\citep{leetaru2014cultural}. The extraction of location data is done through an approach called full text geocoding, developed by Leetaru~\citep{leetaru2012fulltext}. This process applies algorithms to scan news articles and to identify textual mentions of locations using databases of places. Applying the same principle, themes are derived from news items applying extensive topic lists. 
The tone field includes six emotional dimensions. From this field, the average tone of the document is used. This score ranges from -10 (very negative) to +10 (very positive), with zero being neutral~\citep{noauthor_gdelt_2015-1}. The average tone score is based on sentiment mining. This method counts words according to positive and negative pre-compiled dictionaries with the net sentiment representing the overall tone~\citep{hu2004mining}.
The GKG includes c 13,000 themes from c 47,000 sources, from over a billion news articles scanned since 2015. 

\subsection{Predicted variables}
We attempt to predict short-term movements in 10 year BEIR for eight countries -- the US, UK, Germany, Japan, South Africa, Australian, Brazil and Mexico. These countries represent a diverse mix of economies around the world and have mature inflation-linked bond markets.
The BEIR is the difference between nominal and inflation linked bond yields and represents an important indicator of inflation expectations that are priced into the market~\citep{pimco_nodate}.

\subsection{Explanatory variables}
Unlike the majority of research in this field, this study models daily frequency data. We attempt to predict short-term movements in BEIR for eight countries using market-based variables and narrative-based features derived from GDELT. 
Market-based measures of inflation compensation exhibit a strong sensitivity to data surprises, both in terms of changes in market prices and news~\citep{bauer2014inflation}.
We utilise financial variables to capture the impact of financial market conditions such as market participants' perceptions of risk and economic outlook on BEIR~\citep{ciccarelli2009drives}. In addition, we incorporate commodities, as past research has established the presence of clear links between changes in commodity prices and changes in BEIR~\citep{bis_nodate,da2021changes}. Overall, we consider:

\begin{itemize}
  \item Each country's stock market index;
  \item Each country's FX rate;
  \item Each country's yield curve steepener;
  \item Gold price;
  \item Crude oil price;
  \item Bloomberg Commodity Total Return index.
\end{itemize}

In accordance to hypothesis $H_1$, we stipulate that market-based variables account for information that is already priced into BEIR, while variables derived from news narrative contain information not yet reflected in the markets and hence add value to predictions of BEIR movements. Therefore, we include features extracted from GDELT themes and their associated average tone as high-frequency variables to account for data surprises. 

\subsection{GDELT data sets}
\label{sec:gdelt_data_sets}
This section explains the filtering process applied to GDELT data for creating the data sets used in this study. 
In order to identify relevant data surprises as per $H_1$, we attempt to create data sets that reflect the economic and financial environment in terms of GDELT themes for each of the eight respective countries considered in this study.
In order to achieve this, we include news items where the themes field contains at least three themes that are related to economic or financial topics. News are always reported in context, hence we expect themes to also contain other topics besides economic and financial ones. In addition, a country filter is applied to GDELT's GKG location field. 
Applying these filters, we extract the themes and their associated average tone score, and aggregate the data to daily frequency.

For each parsed news article, the themes field is a string of labels including all themes the GDELT algorithm identifies. GDELT themes are very nuanced and we utilise their taxonomy to classify them into categories such as ``environment'', ``health'' or ``political'' (see appendix \ref{sec:gdelt theme categories} for a list of all theme categories). These categories stand for events or conditions, except for four purely descriptive theme groups (``actor'', ``ethnicity'', ``language'', ``point of interest'' and ``animal''), which are therefore omitted. On average, this step reduces the number of unique themes in each respective country data set from roughly 7,000 to roughly 5,000. 

\begin{figure}[pos=h]
	\centering
	\includegraphics[scale=.55]{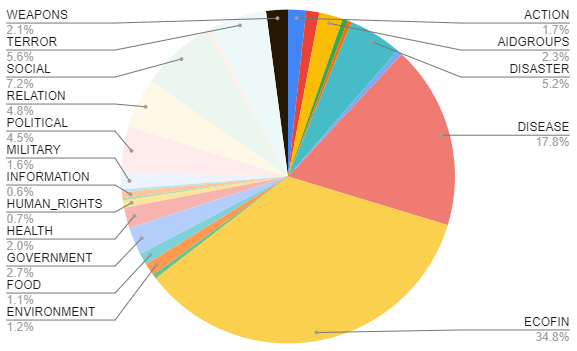}
	\caption{Average split of theme categories}
	\label{FIG:1}
\end{figure}

Figure \ref{FIG:1} illustrates that, on average, economic and financial themes account for around 35\% of all themes, followed by disease and disaster related themes.

\FloatBarrier
\FloatBarrier

\subsection{Data preprocessing}
This section sets out the data preprocessing steps. 

The augmented Dickey Fuller unit root test is applied to the explanatory variables and stationarity is rejected at 5\% significance for all of the market-based variables and most of the GDELT variables. To address the issue of non-stationarity and to reduce noise contained in daily frequency data, we use the five business day difference of these variables for modelling BEIR. Applying the augmented Dickey Fuller test to the differenced data permits us to not reject stationarity for all variables. The variables are then standardized by removing the mean and scaling to unit variance.

We attempt to predict if a country's respective BEIR will increase or decrease over the next five business days. Therefore, we transform the five day change in each country's BEIR into binary variables, i.e., zero (one) for a decrease (increase) in the respective BEIR. Across countries, the two classes are balanced and there is no case when the BEIR did not see a change in levels over five days.

\section{Predictive modelling} 
\label{modelling}

This section describes the analysis that is performed to assess if features derived from news narrative improve predictions of BEIR movements.

Since the predicted variable is binary, this is a classification task. We use five different types of machine learning algorithms to predict BEIR movements one day ahead (see section \ref{lab:model specifications} for model specifications):

\begin{itemize}
    \item{Logistic Regression};
    \item{Support vector classifier};
    \item{Random forest classifier};
    \item{XGBoost classifier};
    \item{Multilayer perceptron classifier}.
\end{itemize}

When modelling differenced data, smaller and larger time intervals such as three, five and ten days have been explored; a difference of five business days yielded the best results across models and predicted variables. Likewise, we have experimented with predictions one, three and five days ahead, with the former generating the best outcomes.

GDELT data sets are high dimensional and exhibit multicollinearity. Therefore, we apply partial least squares (PLS) as dimensionality reduction technique to extract components for inclusion in the above-mentioned models. PLS incorporates information from predicted variable and explanatory variables when calculating scores and loadings, which are chosen to maximise the covariance between predicted variable and explanatory variables \citep{de1993simpls}. First, we predict each country's BEIR with the market-based explanatory variables using a Logistic Regression model. Then, PLS is implemented on the residuals from these predictions. The residuals represent the part of the predicted variable that is not explained and hence, applying PLS to the GDELT data sets adds additional information to the explanatory variables. The orthogonal relationship between the predicted variable and the residuals retains the orthogonality between the PLS components and the market-based explanatory variables.
For each country in our analysis, the first five PLS components explain more than 80\% of the variation in the respective predicted variable. We select five PLS components as in cross-validation analysis, the residual sum of squares is increasing in models with more than five components ~\citep{tobias1995introduction}.

As benchmark, we employ each respective model -- Logistic Regression, Support Vector Classifier, Random Forest Classifier, XGBoost Classifier and Multilayer Perceptron Classifier -- incorporating the market-based explanatory variables only.

Performance is assessed using walk-forward cross-validation, using a five-fold split. This technique is suitable for time series data as in the $k$-th split, it returns the first $k$ folds as train set and the $(k+1)$-st fold as test set.

As performance metrics, precision (the number of true positives divided by the total number of positives), recall (the number of true positives divided by the sum of the number of true positives and the number of false negatives) and the $F_1$ score are used. The $F_1$ score is the harmonic mean of precision and recall, with values ranging from 0 to 1. 

The McNemar test is employed to assess whether model predictions are significantly different from benchmark predictions. This test delivers robust results when determining whether the performance of one classifier is significantly different from another~\citep{dietterich1998approximate}.

\section{Findings}
\label{sec:findings}

This section presents the results from the predictive modelling described in section~\ref{modelling}.

The GDELT data sets are condensed into five PLS components, which are -- together with the market-based explanatory variables -- included into the models set out in section~\ref{modelling} to predict the five day movements of each country's respective BEIR, one day ahead.

In table~\ref{tab:delta_F1}, column names refer to Logistic Regression (LG), support vector classifier (SV), random forest classifier (RF), XGboost classifier (XG) and multilayer perceptron classifier (MLP), respectively.

\FloatBarrier

\begin{center}
\begin{table}[h]
\begin{adjustbox}{width=\columnwidth,center}
\begin{tabular}{|l||*{5}{c|}}\hline
\backslashbox{BEIR for}{Model}&\makebox[4em]{LG}& \makebox[4em]{SV}& \makebox[4em]{RF}&\makebox[4em]{XG}&\makebox[4em]{MLP}\\\hline

US  & \cellcolor{blue!15}0.0947&\cellcolor{blue!15}0.0324& \cellcolor{blue!15}0.1968&\cellcolor{blue!15}0.2158&\cellcolor{blue!15}0.1554\\
UK  & \cellcolor{blue!15}0.1509 & \cellcolor{blue!15}0.0337 & \cellcolor{blue!15}0.1344&\cellcolor{blue!15}0.1590&\cellcolor{blue!15}0.1743\\
Germany  & \cellcolor{blue!15}0.1464 & \cellcolor{blue!15}0.1439& \cellcolor{blue!15}0.2127&\cellcolor{blue!15}0.2097& \cellcolor{blue!15}0.1652\\
Japan  & \cellcolor{blue!15}0.0525& \cellcolor{blue!15}0.0280& \cellcolor{blue!15}0.0717&\cellcolor{blue!15}0.0917&\cellcolor{blue!15}0.0512\\
South Africa  &\cellcolor{blue!15} 0.1986 & \cellcolor{blue!15}0.1862 & \cellcolor{blue!15}0.2578&\cellcolor{blue!15}0.2590&\cellcolor{blue!15}0.2512\\
Australia  & \cellcolor{blue!15}0.0262 & \cellcolor{blue!15}0.0341& \cellcolor{blue!15}0.0813&\cellcolor{blue!15}0.0886&\cellcolor{blue!15}0.0690\\
Brazil  & \cellcolor{blue!15}0.2020 & \cellcolor{blue!15}0.2383 & \cellcolor{blue!15}0.2340 & \cellcolor{blue!15}0.2320 & \cellcolor{blue!15} 0.1935\\
Mexico  & \cellcolor{red!15}-0.2252 & \cellcolor{red!15}-0.1534& \cellcolor{red!15}-0.1260&\cellcolor{red!15}-0.1232&\cellcolor{red!15}-0.1522\\

\hline
\end{tabular}
\caption{Results of the models described in section \ref{modelling} applied to BEIR. Numbers in blue (red) represent the improvement (deterioration) in $F_1$ score compared to the respective benchmark.}
\label{tab:delta_F1}
\end{adjustbox}
\end{table}
\end{center}

For seven out of eight countries, features derived from GDELT themes improve one day ahead predictions of five day BEIR movements. Across countries, the Logistic Regression and XGBoost models achieve the best performance (see section~\ref{sec: raw scores} for Recall, Precision and F1 scores for all respective models and benchmarks).\\ 
The McNemar statistics show that model predictions are statistically different to benchmark predictions (see table \ref{tab:McNemar_statistic}).

\subsection{Feature importance analysis}
\label{sec:interpretability_study}

In this section, we provide insights into the models' performance drivers. For this purpose, we examine the coefficients of the best performing models, Logistic Regression and the XGBoost classifier, for each respective country.

In the Logistic Regression models, the first PLS component represents the largest coefficient in absolute terms, followed by the second and third PLS component, with all PLS components showing a statistical significance of 1\% across countries. 

Similarly, for XGBoost models, the first three components are the three most important performance drivers, according to each model's feature importances. This metric quantifies the gain in accuracy a variable contributes to each tree in the model. A higher value indicates a variable's higher importance for making predictions~\citep{friedman2001elements}.

We analyse the loadings to get insights into the relationship between the first component and the GDELT themes. Loadings represent the strength of relationship between the original sentiment scores for each theme and the PLS components, quantifying the importance of the underlying themes in each of them. 

GDELT themes represent very nuanced events or conditions and are mapped to 25 distinct theme categories using the GDELT naming convention. For example, themes containing the word ``disease'' are mapped to the disease category; themes including the word ``weapon'' are mapped to the weapons category, etc (see section \ref{sec:gdelt theme categories} for detailed list).

\FloatBarrier
\begin{figure}[pos=h]
	\centering
	\includegraphics[scale=.38]{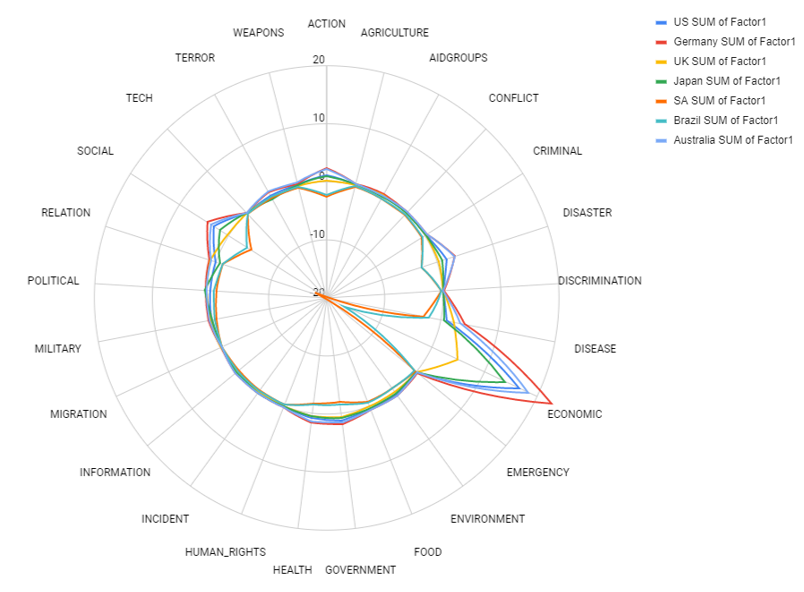}
	\caption{Relationship of first PLS component with theme categories}
	\label{FIG:2}
\end{figure}
\FloatBarrier

The radar chart in Figure \ref{FIG:2} shows the theme categories associated with the loadings of the first PLS components for those Logistic Regression and XGBoost models that outperform their benchmark (see Table \ref{tab:F1_scores}).

Figure \ref{FIG:2} shows that economic and financial themes (``ecofin'') have the strongest relationship with the first PLS component across countries, indicating that economic themes are the main driver of this component and therefore have the strongest predictive power, with some contribution also from the ``social'', ``health'', and ``government'' themes. PLS components 2 and 3 are driven by a mix of theme categories, with no single category standing out as key driver of the respective component (figures are available on request).

\subsection{Cross-country impact of narrative on BEIR}

In this section, we investigate the cross-country impact of narrative on BEIR. We estimate this effect by utilising Graphical Granger Causality (GGC) modelling \citep{lozano2009grouped}. 

Granger Causality tests pairs of features to establish if there are statistical relationships between those variables~\citep{granger1969investigating}. It is important to note that Granger causality is testing for precedence rather than true causality and may be found even in the absence of an actual causal connection~\citep{leamer_1985}. Following Arnold \emph{et al}, we apply the ``exhaustive graphical Granger method''~\citep{arnold2007temporal} to every variable pair, where we use the five day difference in all 74 variables and a lag of one day. For this test, the null hypothesis  states that a lagged variable does not Granger-cause a variable at a significance level of 5\%, while the alternative hypothesis states that the lagged variable Granger-causes a variable at the same significance level. Since we conduct multiple testing, we employ the Benjamini-Hochberg (BH) procedure to correct for false positives~\citep{benjamini2005false}.

\begin{figure*}
  \includegraphics[width=\textwidth]{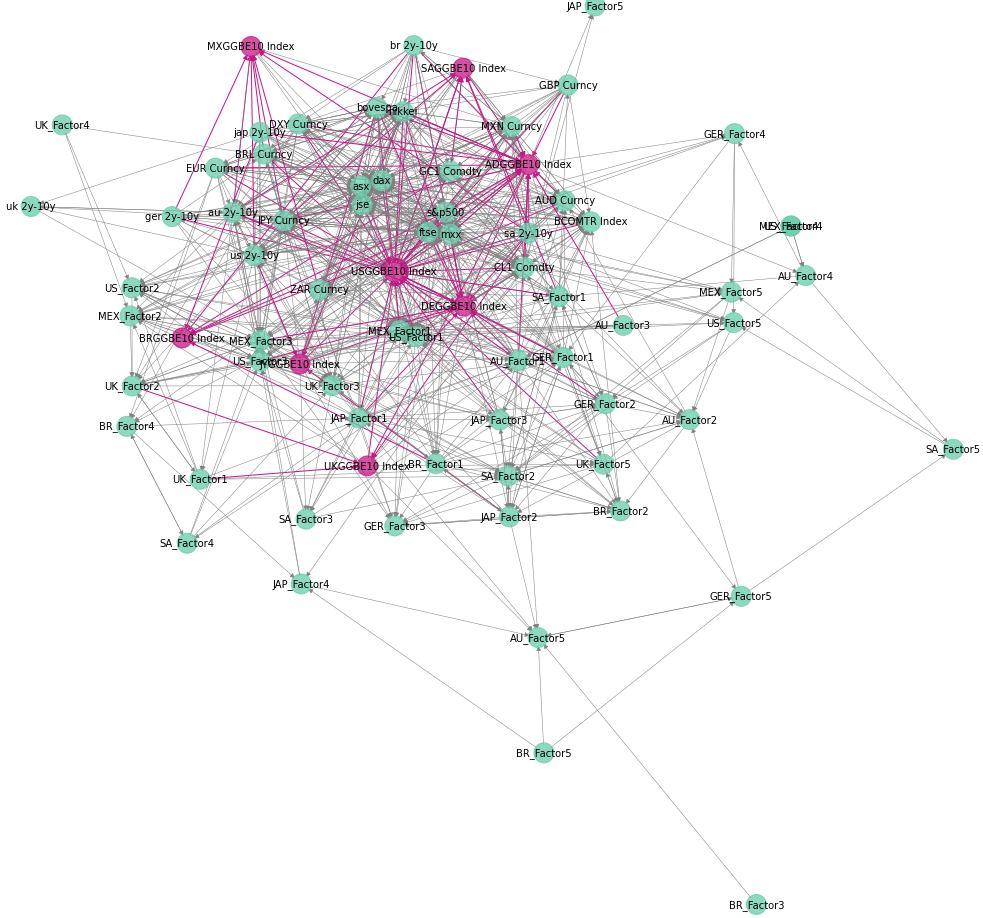}
  \caption{Graphical Granger Causality between variables}
  \label{FIG:3}
\end{figure*}

If two variables have a statistically significant adjusted $p$-value against the aforementioned null hypothesis, the variable pair forms a directed edge where the first variable represents the source node. Since all variables have at least one in- or outbound connection, all variables are preserved as nodes. Based on this analysis, we construct a directed knowledge graph containing 74 nodes and 817 edges (see figure~\ref{FIG:3}), resulting in a connected graph with a density of 0.1512. 
For each of the eight countries, the five PLS components extracted from news themes and their associated tone are referred to as narrative-based factors.

In order to gain an understanding of the role and importance of news narrative in this knowledge graph, we calculate the inbetweenness centrality scores for all nodes~\citep{friedman2001elements}. According to this metric, nodes representing either the first, second or third PLS component for the eight respective countries rank in the top quintile, supporting our findings from section~\ref{sec:interpretability_study}, where we established that the first three PLS components have the strongest predictive power across countries. Conversely, the least central nodes correspond to features with low predictive power (both in terms of coefficient size and feature importance). 

To identify the cross-country effect of narrative on BEIR, we examine the predecessor nodes for each country's BEIR. Seven BEIR have incoming edges from market-based explanatory variables as well as PLS components. Mexico only has incoming edges from market-based variables, which is consistent with our results from \ref{sec:findings} that news narrative does not significantly impact the country's BEIR.

While all BEIRs have incoming edges from local PLS components, only the US and Germany have incoming edges from other countries' narrative-based factors. For instance, the US BEIR is impacted by PLS components relating to  Germany, Japan, South Africa and Australia. The German BEIR has incoming edges from PLS components associated with the US, Japan and Australia.  

The findings suggest that our approach identifies so-called spillover effects from news narrative on BEIR only for two of the eight economies we examined, the US and Germany. 

The incoming edges for all BEIR suggest that cross-country spillover effects are also captured by the market-based explanatory variables in the case of other countries, reflecting the global interconnectedness of local capital markets.\\
Figure~\ref{FIG:3} shows the knowledge graph, with BEIR and their incoming edges highlighted in pink.

\FloatBarrier
\FloatBarrier

\begin{figure*}
\begin{center}
\includegraphics[width=0.65\textwidth]{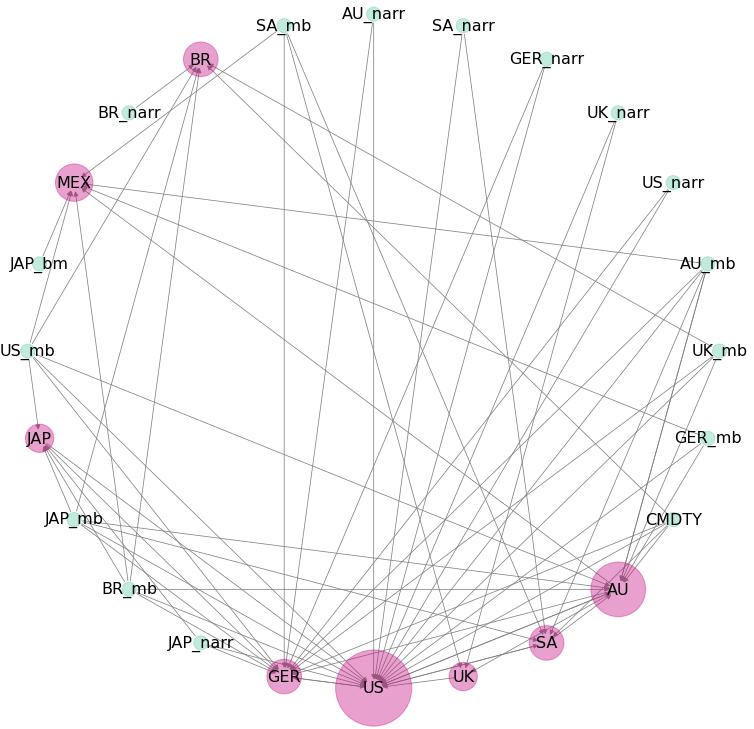}
\end{center}
  \caption{Graphical Granger Causality: incoming edges of BEIR}
  \label{FIG:4}
\end{figure*}

For clarity purposes, figure \ref{FIG:4} only shows each country's BEIR and their respective incoming edges. Nodes with country names refer to the respective country's BEIR (pink nodes), the suffix ``\_mb'' summarises nodes representing country-specific market-based explanatory variables, the node ``CMDTY'' stands for commodity-related variables and the suffix ``\_narr'' describes country-specific media narratives. The node size accounts for the number of incoming edges.

\FloatBarrier
\FloatBarrier

Our analysis highlights the importance of both news narrative for fluctuations in a country's BEIR, with cross-country spillovers of news narrative observed in movements of the US and German BEIR. The findings suggest that variables derived from news narrative capture the market's ``animal spirits''~\citep{keynes1936general,shiller2017narrative,tuckett2014bringing}.

\section{Discussion}

In this study, we propose an effective big data filtering method to condense large volumes of data and to create data sets representing economic and financial news for eight different countries. We predict the five day movements in BEIR for these countries one day ahead using market-based explanatory variables and narrative-based features, calibrating five types of machine learning models. We find that the models generally outperform corresponding benchmarks that do not include such features, with Logistic Regression and XGBoost models beating their respective benchmarks for seven out of eight countries. We complement those results with a feature importance analysis, illustrating that economic and financial themes are the key drivers of the first PLS component, which is the most important variable in Logistic Regression and XGBoost models. Based on these findings, we claim that narrative-based features significantly improve predictions of movements in market-based inflation expectations, and therefore cannot reject $H_1$.

We examine the cross-country impact of narrative on BEIR via Graphical Granger Causality (GGC) modelling. Our results indicate that cross-country spillover effects of narrative exist for US and German market-based inflation expectations. Mexico is only influenced by market-based variables and the remaining BEIR considered are impacted by local narrative only. Therefore, we cannot reject $H_2$ in the case of the US and Germany, and reject it for the other countries considered in our work.

This study advances existing research on the prediction of market-based inflation expectations in at least three aspects. First, we demonstrate how themes and their associated sentiment from news narrative can be incorporated into machine learning models to improve predictions of BEIR movements for a range of diverse economies. Second, our work contributes to literature on forecasting market-based inflation expectations by presenting a data-driven approach to operationalise concepts such as information extraction, feature generation and predictive modelling. Our feature importance analysis enhances the interpretability of our results, showing that economic and financial themes have the strongest predictive power, with additional secondary contributions from themes related to health and government. Third, we contribute to work on the cross-country impact of news narrative on BEIR leveraging GGC and graph analytics.

Before concluding, we should acknowledge some potential limitations of our study. First, our study is a proof of concept and does not attempt to fully optimise performance. The models we use only employ a limited selection of market-based explanatory variables as controls and real-world applicability could be enhanced by expanding to a larger range of explanatory variables. Second, GDELT data has a relatively short track record, starting end of February 2015. There are plans to add historical data going back to 1979 and it will be interesting to rerun the experiment once a longer track record is in place.

\section{Conclusions}
Our research presents a novel approach to incorporating themes and sentiment from global newspapers into the predictions of market-based inflation expectations movements. Our findings show that features derived from narrative significantly improve predictions of BEIR movements for seven out of eight countries considered, with economic and financial themes being the key performance drivers. We identify cross-country spillovers effects of news narrative on BEIR in the US and Germany, while five further BEIR considered are impacted by local news flow only.

\section{Appendix}

\subsection{GDELT theme categories}
\label{sec:gdelt theme categories}

All GDELT themes are mapped to one of 30 theme categories. For example, the ``Weapons'' contains 81 themes such as \texttt{``TAX\_WEAPONS\_GUNS''}, \texttt{``TAX\_WEAPONS\_BOMB''} and\\ \texttt{``TAX\_WEAPONS\_SUICIDE\_BOMB''}. We removed themes belonging to ``Actor'', ``Language'', ``Animal'' ``Points of Interest'' and ``Ethnicity'' for our analysis as they are purely descriptive. Theme categories are:

\begin{itemize}
  \item Ecofin
  \item Disease
  \item Actor
  \item Action
  \item Language
  \item Ethnicity
  \item Animal
  \item Disaster
  \item Social
  \item Relation
  \item Political
  \item Health
  \item Weapons
  \item Military
  \item Terror
  \item Environment
  \item Food
  \item Government
  \item Aid groups
  \item Information
  \item Conflict
  \item Emergency
  \item Human rights
  \item Migration
  \item Agriculture
  \item Discrimination
  \item Incident
  \item Criminal
  \item Tech
  \item Points of interest
\end{itemize}

\subsection{Model specifications}
\label{lab:model specifications}
This section provides details on the algorithms that were employed for modelling BEIR. Classes are balanced.

\subsubsection{Logistic Regression}
The Logistic Regression model uses the logistic function to squeeze the output of a linear equation between 0 and 1, solving

\begin{align*}
P = \frac{1}{1 + e^{-(\beta_0 + \beta_1x_1 + ... + \beta_p x_p)}}
\end{align*}
for the coefficients $\beta_i$. We use the $L_2$ penalty and the limited-memory Broyden-Fletcher-Goldfarb-Shanno (lbfg) solver \citep{fletcher2013practical}.

\subsubsection{Support Vector Classifier}
The Support Vector Classifier takes in separable observations and generates a hyperplane that best separates the two classes, creating a decision boundary that maximises the margins for both classes. The algorithm solves

\[\min_{w \in \Re^{d}, \xi \in \Re^{+}} = \frac{1}{2}||w||^{2} + C \sum_{i}^N \xi_i \]
where $w$ are the parameters, $C$ is a hyperparameter controlling the penalty for misclassification and $\xi_i$ is a slack variable. We set $C$ at 1.0 and use a radial basis function as kernel.


\subsubsection{Random Forest Classifier}
We use a bootstrapped Random Forest Classifier with 50 trees and a minimum sample split of 2.

\subsubsection{XGBoost Classifier}
We use an XGBoost Classifier with a maximum tree depth of 10 for base learners.

\subsubsection{Multilayer Perceptron Classifier}
We use a Multilayer Perceptron Classifier that optimizes the log-loss function using lbfgs \citep{fletcher2013practical}4. We use 10 hidden layers, relu as activation function, adam as solver and an $L_2$ penalty of $10^{-4}$. 

\subsection{F1, Recall and Precision scores}
\label{sec: raw scores}

In the following tables, the column names refer to Logistic Regression, Support Vector classifier, Random Forest classifier, XGBoost classifier and Multilayer Perceptron classifier, respectively.

Table \ref{tab:F1_scores} contains the $F_1$ scores for all models and countries.

\begin{center}
\begin{table}[h]
\begin{adjustbox}{width=\columnwidth,center}
\begin{tabular}{|l||*{5}{c|}}\hline
\backslashbox{BEIR for}{Model}&\makebox[4em]{LG}& \makebox[4em]{SV}& \makebox[4em]{RF}&\makebox[4em]{XG}&\makebox[4em]{MLP}\\\hline

US  & 0.8124&0.7427& 0.8215&0.8400&0.8160\\

UK  & 0.7174 & 0.5726 & 0.6552 &0.6800&0.7166\\

Germany  & 0.7096 & 0.6567& 0.7347&0.7226& 0.7023\\

Japan  & 0.5582& 0.5455& 0.5417&0.5644&0.5793\\

South Africa  & 0.8300 & 0.8289 & 0.8704 &0.8723&0.8741\\

Australia  & 0.6533 & 0.6569& 0.6492&0.6512&0.6629\\

Brazil  & 0.6627 & 0.5403 & 0.7074 & 0.6947 & 0.6676\\

Mexico  & 0.3835 & 0.4483& 0.4196& 0.4216&0.4074\\

\hline
\end{tabular}
\caption{Results of the models described in section \ref{modelling} applied to BEIR. Numbers represent the $F_1$ score.}
\label{tab:F1_scores}
\end{adjustbox}
\end{table}
\end{center}

Table \ref{tab:Recall_scores} contains the Recall scores for all models and countries.

\begin{center}
\begin{table}[h]
\begin{adjustbox}{width=\columnwidth,center}
\begin{tabular}{|l||*{5}{c|}}\hline
\backslashbox{BEIR for}{Model}
&\makebox[4em]{LG}& \makebox[4em]{SV}& \makebox[4em]{RF}&\makebox[4em]{XG}&\makebox[4em]{MLP}\\\hline

US & 0.8064 & 0.7551&0.7829& 0.8138&0.8049\\

UK & 0.7047 & 0.5568 & 0.6450 & 0.6786 &0.7057\\

Germany & 0.7016 & 0.6386 & 0.7474& 0.7412&0.7060\\

Japan & 0.5537 & 0.5260& 0.5409& 0.5761&0.5932\\

South Africa & 0.8855 &0.8184 & 0.8727 & 0.8771 &0.8872\\

Australia & 0.6468 & 0.6513 & 0.6523& 0.6475&0.6564\\

Brazil & 0.7122& 0.5257 & 0.7465 & 0.7566 & 0.7432 \\

Mexico & 0.3846 & 0.4600 & 0.4306& 0.4568& 0.4111\\

\hline
\end{tabular}
\caption{Results of the models described in \ref{modelling} applied to BEIR. Numbers represent the Recall score.}
\label{tab:Recall_scores}
\end{adjustbox}
\end{table}
\end{center}

Table \ref{tab:Precision_score} contains the Precision scores for all models and countries.

\begin{center}
\begin{table}[h]
\begin{adjustbox}{width=\columnwidth,center}
\begin{tabular}{|l||*{5}{c|}}\hline
\backslashbox{BEIR for}{Model}
&\makebox[4em]{LG}& \makebox[4em]{SV}& \makebox[4em]{RF}&\makebox[4em]{XG}&\makebox[4em]{MLP}\\\hline

US & 0.8246 & 0.7383&0.8721& 0.8731&0.8330\\

UK & 0.7328 & 0.5920 & 0.6747 & 0.6851 &0.7301\\

Germany & 0.7250 & 0.6884 & 0.7297& 0.7130&0.7069\\

Japan & 0.5751 & 0.5834& 0.5589& 0.5678&0.5797\\

South Africa & 0.8746 &0.8480 & 0.8687 & 0.8686 &0.8617\\

Australia & 0.6682 & 0.6672 & 0.6501& 0.6600&0.6742\\

Brazil & 0.6208 & 0.5732 & 0.6741 & 0.6454 & 0.6080 \\

Mexico & 0.3837 & 0.4408 & 0.4101& 0.4084& 0.4049\\

\hline
\end{tabular}
\caption{Results of the models described in \ref{modelling} applied to BEIR. Numbers represent the Precision score.}
\label{tab:Precision_score}
\end{adjustbox}
\end{table}
\end{center}

Table \ref{tab:BM_F1_scores} contains the $F_1$ scores for all benchmark models and countries.

\begin{center}
\begin{table}[H]
\begin{adjustbox}{width=\columnwidth,center}
\begin{tabular}{|l||*{5}{c|}}\hline
\backslashbox{BEIR for}{Model}
&\makebox[4em]{LG}& \makebox[4em]{SV}& \makebox[4em]{RF}&\makebox[4em]{XG}&\makebox[4em]{MLP}\\\hline

US & 0.7177 & 0.7103&0.6247& 0.6242&0.6606\\

UK & 0.5765 & 0.5389 & 0.5208 & 0.5210 &0.5423\\

Germany & 0.5632 & 0.5128 & 0.5230& 0.5129&0.5371\\

Japan & 0.5057 & 0.5175& 0.4700& 0.4727&0.5281\\

South Africa & 0.6314 &0.6427 & 0.6126 & 0.6133 &0.6229\\

Australia & 0.6271 & 0.6228 & 0.5679& 0.5626&0.5939\\

Brazil & 0.4607 & 0.3020 & 0.4734 & 0.4627 & 0.4741\\

Mexico & 0.6087 & 0.6017 & 0.5456& 0.5448& 0.5596\\

\hline
\end{tabular}
\caption{Results of the models described in \ref{modelling} applied to BEIR. Numbers represent the benchmark $F_1$ score.}
\label{tab:BM_F1_scores}
\end{adjustbox}
\end{table}
\end{center}

\FloatBarrier

Table \ref{tab:BM_Recall_scores} contains the Recall scores for all benchmark models and countries.

\FloatBarrier

\begin{center}
\begin{table}[h]
\begin{adjustbox}{width=\columnwidth,center}
\begin{tabular}{|l||*{5}{c|}}\hline
\backslashbox{BEIR for}{Model}
&\makebox[4em]{LG}& \makebox[4em]{SV}& \makebox[4em]{RF}&\makebox[4em]{XG}&\makebox[4em]{MLP}\\\hline

US & 0.6685 & 0.8237&0.6422& 0.6411&0.7156\\

UK & 0.5923 & 0.5057 & 0.4931 & 0.5090 &0.5085\\

Germany & 0.6188 & 0.4383 & 0.5137& 0.5039&0.5074\\

Japan & 0.4925 & 0.4987& 0.4674& 0.4718&0.5188\\

South Africa & 0.6549 &0.6145 & 0.5992 & 0.6177 &0.5986\\

Australia & 0.6411 & 0.6542 & 0.5729& 0.5720&0.6123\\

Brazil & 0.5507 & 0.2282 & 0.4751 & 0.4703 & 0.4369\\

Mexico & 0.6000 & 0.5873 & 0.5399& 0.5414& 0.5257\\

\hline
\end{tabular}
\caption{Results of the models described in \ref{modelling} applied to BEIR. Numbers represent the benchmark Recall score.}
\label{tab:BM_Recall_scores}
\end{adjustbox}
\end{table}
\end{center}

\FloatBarrier

Table \ref{tab:BM_Precision_scores} contains the Precision scores for all benchmark models and countries.

\FloatBarrier

\begin{center}
\begin{table}[h]
\begin{adjustbox}{width=\columnwidth,center}
\begin{tabular}{|l||*{5}{c|}}\hline
\backslashbox{BEIR for}{Model}
&\makebox[4em]{LG}& \makebox[4em]{SV}& \makebox[4em]{RF}&\makebox[4em]{XG}&\makebox[4em]{MLP}\\\hline

US & 0.7847 & 0.6320&0.6171& 0.6159&0.6295\\

UK & 0.5631 & 0.5837 & 0.5556 & 0.5370 &0.5855\\

Germany & 0.5213 & 0.6339 & 0.5416& 0.5283&0.5812\\

Japan & 0.5223 & 0.5409& 0.4750& 0.4784&0.5401\\

South Africa & 0.6180 &0.6803 & 0.6363 & 0.6232 &0.6661\\

Australia & 0.6243 & 0.6000 & 0.5711& 0.5624&0.5827\\

Brazil & 0.4298 & 0.5823 & 0.4767 & 0.4608 & 0.5294\\

Mexico & 0.6213 & 0.6219 & 0.5564 & 0.5528 & 0.6058\\

\hline
\end{tabular}
\caption{Results of the models described in \ref{modelling} applied to BEIR. Numbers represent the benchmark Precision score.}
\label{tab:BM_Precision_scores}
\end{adjustbox}
\end{table}
\end{center}

\FloatBarrier

\subsection{McNemar statistics}

Table \ref{tab:McNemar_statistic} contains the McNemar statistics for all classifiers, illustrating that all model predictions are statistically different to the benchmark predictions at either 5\% or 1\% significance. 

The columns names refer to Logistic Regression, Support Vector classifier, Random Forest classifier, XGBoost classifier and Multilayer Perceptron classifier, respectively.

\FloatBarrier

\begin{center}
\begin{table}[h]
\begin{adjustbox}{width=\columnwidth,center}
\begin{tabular}{|l||*{5}{c|}}\hline
\backslashbox{BEIR for}{Model}
&\makebox[4em]{LG}& \makebox[4em]{SV}& \makebox[4em]{RF}&\makebox[4em]{XG}&\makebox[4em]{MLP}\\\hline

US & 0.0264 & 0.0253 & 0.0328 & 0.0322 & 0.0303\\

UK  & 0.0236 & 0.0246 & 0.0191 & 0.0213 & 0.0242\\

Germany  & 0.0192 & 0.0222 & 0.0266 & 0.0252 & 0.0218\\

Japan & 0.0300 & 0.0295 & 0.0309 & 0.0303 & 0.0282\\

South Africa  & 0.0356 & 0.0367 & 0.0386 & 0.0355 & 0.0352\\

Australia & 0.0107 & 0.0158 & 0.0160 & 0.0167 & 0.0156\\

Brazil & 0.0149 & 0.0208 & 0.0215 & 0.0243 & 0.0175\\

Mexico & 0.0231 & 0.0235 & 0.0203 & 0.0194 & 0.0226\\

\hline
\end{tabular}
\caption{Numbers represent McNemar statistics, comparing model to benchmark predictions.}
\label{tab:McNemar_statistic}
\end{adjustbox}
\end{table}
\end{center}

\FloatBarrier

\section*{Declaration of competing interest}
The authors declare that they have no known competing financial interests or personal relationships that could have influenced the work reported in this paper.

\bibliographystyle{apacite}

\bibliography{cas-refs}

\end{document}